# Atomic systems with bound states of fermions in the Schwarzschild, Reissner-Nordström fields as candidates for the role of dark matters particles

V.P.Neznamov[*], I.I.Safronov, V.E.Shemarulin

*FSUE "RFNC-VNIIEF"*
*37 Mira pr., Sarov, Nizhny Novgorod region, 607188, Russia*

Abstract

After transition from the Dirac equation to the Schrödinger-type relativistic equation with effective potentials of the Schwarzschild and Reissner-Nordström (RN) fields, the existence of the stationary state of fermions with real square-integrable radial wave functions is proved. The fermions are localized near the event horizon within the range from zero to several fractions or a few units of the Compton wavelength of a fermion as a function of the gravitational and electromagnetic coupling constants and the angular and orbital momenta $j,l$. Electrically neutral atomic-type systems (Schwarzschild and RN collapsars with fermions in bound states) are proposed as particles of dark matter.

---

[*] vpneznamov@vniief.ru
   vpneznamov@mail.ru

## 1. Introduction

Earlier, in [1], [2] the existence of the stationary bound states of fermions localized near the event horizons, in the intervals from zero to fractions or a few units of the Compton wavelength of fermions as a function of the gravitational and electromagnetic coupling constants and the angular and orbital momenta $j,l$ was proved for Schwarzschild and Reissner-Nordström metrics using the relativistic equations of Schrödinger type with effective potentials.

In [1], the energies of bound stationary states of fermions for the fields with rotation (Kerr and Kerr-Newman space-time) were also announced.

In this paper, the results obtained in [1], [2] are briefly reproduced to form the basis of the suggestion that collapsars with fermions in bound states can be considered as dark matter particles.

The following system of units $\hbar = c = 1$ is mostly used in this paper

$$g_{\alpha\beta} = \text{diag}[1,-1,-1,-1]. \tag{1}$$

The following notations are used:

- The Schwarzschild field with point mass $M$ : $r_0 = 2GM/c^2$ is the radius of the event horizon, dimensionless variables $\rho = r/l_c$, $r_0/l_c = 2\alpha$, $\alpha = GMm/\hbar c = Mm/M_P^2$, $M_P$ is the Plank mass, $l_c = \hbar/mc$ is the Compton wavelength of the fermion, $m, E$ are the mass and energy of the Dirac particle, $\varepsilon = E/m$;

- The Reissner-Nordström field with mass $M$ and charge $Q$:
$r_Q = \sqrt{G}|Q|/c^2$, $\rho = r/l_c$, $\alpha_Q = r_Q/l_c = \sqrt{G}QM/\hbar c$, $\alpha_{em} = eQ/\hbar c$, $e$ is the electric charge of the fermion; $\rho_\pm = \alpha \pm \sqrt{\alpha^2 - \alpha_Q^2}$, $\rho_\pm$ are dimensionless radii of exterior and interior event horizons at $\alpha^2 > \alpha_Q^2$;

- The Kerr field with mass $M$ and momentum $J$, and the Kerr-Newman fields with mass $M$, momentum $J$ and charge $Q$: $a = J/Mc$, $\rho = r/l_c$, $\alpha_a = a/l_c$, $\rho_\pm = \alpha \pm \sqrt{\alpha^2 - \alpha_a^2 - \alpha_Q^2}$, $\rho_\pm$ are dimensionless radii of exterior and interior event horizons at $\alpha^2 > \alpha_a^2 + \alpha_Q^2$.

## 2. Is it possible to create conditions of existence of stationary bound states of fermions in external gravitational fields?

The Reissner-Nordström and Schwarzschild metrics are

$$ds^2 = f_{R-N}dt^2 - \frac{dr^2}{f_{R-N}} - r^2\left(d\theta^2 + \sin^2\theta d\varphi^2\right), \tag{2}$$

$$g_{00}^{R-N} = f_{R-N} = \left(1 - \frac{r_0}{r} + \frac{r_Q^2}{r^2}\right) = \left(1 - \frac{2\alpha}{\rho} + \frac{\alpha_Q^2}{\rho^2}\right). \tag{3}$$

At $r_Q = 0$, the Schwarzschild metric is applied

$$g_{00}^s = f_s = 1 - \frac{r_0}{r} = 1 - \frac{2\alpha}{\rho}. \tag{4}$$

The Dirac equation in the Hamiltonian form is

$$i\frac{\partial \Psi_\eta}{\partial t} = H_\eta \Psi_\eta. \tag{5}$$

What do we have?

**1.** Self-adjoint Hamiltonian

$$H_\eta = H_\eta^+ = \sqrt{f_{R-N}}\, m\gamma^0 - i\gamma^0\gamma^3 \left(f_{R-N}\frac{\partial}{\partial r} + \frac{1}{r} - \frac{r_0}{2r^2}\right) -$$
$$-i\sqrt{f_{R-N}}\frac{1}{r}\left[\gamma^0\gamma^1\left(\frac{\partial}{\partial \theta} + \frac{1}{2}\mathrm{ctg}\,\theta\right) + \gamma^0\gamma^2 \frac{1}{\sin\theta}\frac{\partial}{\partial \varphi}\right] + \frac{eQ}{r}. \tag{6}$$

**2.** Possibility to separate variables

$$\Psi_\eta = \begin{pmatrix} F(r)\xi(\theta) \\ -iG(r)\sigma^3\xi(\theta) \end{pmatrix} e^{-iEt} e^{im_\varphi \varphi}. \tag{7}$$

$$\xi(\theta) = \begin{pmatrix} {}_{-1/2}Y_{jm_\varphi}(\theta) \\ {}_{1/2}Y_{jm_\varphi}(\theta) \end{pmatrix} = (-1)^{m_\varphi + 1/2} \sqrt{\frac{1}{4\pi}\frac{(j-m_\varphi)!}{(j+m_\varphi)!}} \begin{pmatrix} \cos\theta/2 & \sin\theta/2 \\ -\sin\theta/2 & \cos\theta/2 \end{pmatrix} \begin{pmatrix} \left(\kappa - m_\varphi + \frac{1}{2}\right) P_l^{m_\varphi - 1/2}(\theta) \\ P_l^{m_\varphi + 1/2}(\theta) \end{pmatrix}. \tag{8}$$

**3.** The density of the current of Dirac particles

$$j^0 = \Psi_\eta^+ \Psi_\eta = \left(F(\rho)F^*(\rho) + G(\rho)G^*(\rho)\right)\xi^+(\theta)\xi(\theta), \tag{9}$$

$$j^\rho = \Psi_\eta^+ f_{R-N}\gamma^0\gamma^3 \Psi_\eta = -if_{R-N}\left(F^*(\rho)G(\rho) - F(\rho)G^*(\rho)\right)\xi^+(\theta)\xi(\theta), \tag{10}$$

$$j^\theta = \Psi_\eta^+ \frac{f_{R-N}^{1/2}}{\rho}\gamma^0\gamma^1 \Psi_\eta = -\frac{f_{R-N}^{1/2}}{\rho}\left(F^*(\rho)G(\rho) + F(\rho)G^*(\rho)\right)\xi^+(\theta)\sigma^2\xi(\theta), \tag{11}$$

$$j^\varphi = \Psi_\eta^+ \frac{f_{R-N}^{1/2}}{\rho\sin\theta}\gamma^0\gamma^2 \Psi_\eta = \frac{f_{R-N}^{1/2}}{\rho\sin\theta}\left(F^*(\rho)G(\rho) + F(\rho)G^*(\rho)\right)\xi^+(\theta)\sigma^1\xi(\theta). \tag{12}$$

**3.1.** In general, the current density $j^\rho$ can be nonzero for complex radial functions. In this case, the Hamiltonian is non-Hermitian $(\Phi, H_\eta\Psi) \neq (H_\eta\Phi, \Psi)$. There can be only quasi-stationary states of fermions that decay with time.

**3.2.** For real functions $\left(F^* = F, G^* = G\right)$, the density of the radial current is zero across the entire domain of the wave functions, and the Dirac Hamiltonian is Hermitian.

Current density $j^\theta$ is zero for both complex and real $F(\rho)$ and $G(\rho)$, since $\xi^+(\theta)\sigma^2\xi(\theta) = 0$. On the contrary, current density $j^\varphi$ is nonzero for any of functions $F(\rho)$ or $G(\rho)$.

**4.** IMPORTANT: We limit our consideration to the class of real radial functions $F(\rho), G(\rho)$.

**5.** The system of equations for real radial functions

$$f_{R-N}\frac{dF(\rho)}{d\rho} + \left(\frac{1+\kappa\sqrt{f_{R-N}}}{\rho} - \frac{\alpha}{\rho^2}\right)F(\rho) - \left(\varepsilon - \frac{\alpha_{em}}{\rho} + \sqrt{f_{R-N}}\right)G(\rho) = 0,$$

$$f_{R-N}\frac{dG(\rho)}{d\rho} + \left(\frac{1-\kappa\sqrt{f_{R-N}}}{\rho} - \frac{\alpha}{\rho^2}\right)G(\rho) + \left(\varepsilon - \frac{\alpha_{em}}{\rho} - \sqrt{f_{R-N}}\right)F(\rho) = 0.$$

(13)

The system of equations is real if $f_{R-N} \geq 0$, i.e., intervals $\rho \in [\rho_+, \infty)$, $\rho \in (0, \rho_-]$ are the domains of functions $F(\rho), G(\rho)$.

For the Schwarzschild field: condition $f_s \geq 0$ implies the domain $\rho \in [2\alpha, \infty)$.

**6.** Asymptotics of solutions near the event horizons

**6.1.**
$$F\big|_{\rho \to \rho_+} = \frac{A}{\sqrt{\rho - \rho_+}} \sin\left(\frac{\rho_+^2}{\rho_+ - \rho_-}\left(\varepsilon - \frac{\alpha_{em}}{\rho_+}\right)\ln(\rho - \rho_+) + \varphi_+\right),$$

$$G\big|_{\rho \to \rho_+} = \frac{A}{\sqrt{\rho - \rho_+}} \cos\left(\frac{\rho_+^2}{\rho_+ - \rho_-}\left(\varepsilon - \frac{\alpha_{em}}{\rho_+}\right)\ln(\rho - \rho_+) + \varphi_+\right).$$

(14)

**6.2.**
$$F\big|_{\rho \to \rho_-} = -\frac{B}{\sqrt{\rho_- - \rho}} \sin\left(\frac{\rho_-^2}{\rho_+ - \rho_-}\left(\varepsilon - \frac{\alpha_{em}}{\rho_-}\right)\ln(\rho_- - \rho) + \varphi_-\right),$$

$$G\big|_{\rho \to \rho_-} = \frac{B}{\sqrt{\rho_- - \rho}} \cos\left(\frac{\rho_-^2}{\rho_+ - \rho_-}\left(\varepsilon - \frac{\alpha_{em}}{\rho_-}\right)\ln(\rho_- - \rho) + \varphi_-\right).$$

(15)

**6.3.** For the Schwarzschild field

$$F\big|_{\rho \to 2\alpha} = \frac{A}{\sqrt{\rho - 2\alpha}} \sin\left(2\alpha\varepsilon \ln(\rho - 2\alpha) + \varphi\right),$$

$$G\big|_{\rho \to 2\alpha} = \frac{A}{\sqrt{\rho - 2\alpha}} \cos\left(2\alpha\varepsilon \ln(\rho - 2\alpha) + \varphi\right).$$

(16)

Unpleasant statement:

- in the vicinity of vent horizons, functions $F(\rho), G(\rho)$ are square nonintegrable (for example, normalizing integral $N = \int_{\rho_+}^{\infty} \left( F(\rho)^2 + G(\rho)^2 \right) \rho^2 d\rho$ is logarithmically divergent);

- for $\varepsilon \neq \alpha_{em}/\rho_+$, $\varepsilon \neq \alpha_{em}/\rho_-$ (the Reissner-Nordström field), $\varepsilon \neq 0$ (the Schwarzschild field), the displayed asymptotics show that particles "fall" onto the event horizons.

Positive moment:

- for $\varepsilon = \alpha_{em}/\rho_+$, $\varepsilon = \alpha_{em}/\rho_-$ (the Reissner-Nordström field);

$\varepsilon = 0$ (the Schwarzschild field) asymptotics show the absence of the mode of fermion "fall" onto the event horizons.

In order to admit the existence of stationary bound states of fermions, it is necessary to solve the problem of square nonintegrability of wave functions.

**7.** The Schrödinger type equation with an effective potentials

We transform the system of Dirac equations for radial functions $F(\rho), G(\rho)$ to relativistic equations of Schrödinger type for function $\psi_F(\rho)$ proportional to $F(\rho)$, and for function $\psi_G(\rho)$ proportional to $G(\rho)$.

$$\psi_F(\rho) = F(\rho)\exp\left(\frac{1}{2}\int^{\rho} A_F(\rho')d\rho'\right), \quad \psi_G(\rho) = G(\rho)\exp\left(\frac{1}{2}\int^{\rho} A_G(\rho')d\rho'\right), \quad (17)$$

where $A_F(\rho) = -\frac{1}{B}\frac{dB}{d\rho} - A - D$, $A_G(\rho) = -\frac{1}{C}\frac{dC}{d\rho} - A - D$ and

$$A(\rho) = -\frac{1}{f_{R-N}}\left(\frac{1+\kappa\sqrt{f_{R-N}}}{\rho} - \frac{\alpha}{\rho^2}\right), \quad B(\rho) = \frac{1}{f_{R-N}}\left(\varepsilon + \sqrt{f_{R-N}}\right),$$

$$C(\rho) = -\frac{1}{f_{R-N}}\left(\varepsilon - \sqrt{f_{R-N}}\right), \quad D(\rho) = -\frac{1}{f_{R-N}}\left(\frac{1-\kappa\sqrt{f_{R-N}}}{\rho} - \frac{\alpha}{\rho^2}\right). \quad (18)$$

Equations for $\psi_F(\rho)$ and $\psi_G(\rho)$ are similar to the Schrödinger equation

$$\frac{d^2\psi_F(\rho)}{d\rho^2} + 2\left(E_{Schr} - U_{eff}^F(\rho)\right)\psi_F(\rho) = 0, \quad (19)$$

$$\frac{d^2\psi_G(\rho)}{d\rho^2} + 2\left(E_{Schr} - U_{eff}^G(\rho)\right)\psi_G(\rho) = 0, \quad (20)$$

where

$$E_{Schr} = \frac{1}{2}\left(\varepsilon^2 - 1\right), \quad (21)$$

$$U_{eff}^{F}(\rho) = E_{Schr} + \frac{3}{8}\frac{1}{B^2}\left(\frac{dB}{d\rho}\right)^2 - \frac{1}{4}\frac{1}{B}\frac{d^2B}{d\rho^2} + \frac{1}{4}\frac{d}{d\rho}(A-D) -$$
$$-\frac{1}{4}\frac{(A-D)}{B}\frac{dB}{d\rho} + \frac{1}{8}(A-D)^2 + \frac{1}{2}BC. \tag{22}$$

$$U_{eff}^{G}(\rho) = E_{Schr} + \frac{3}{8}\frac{1}{C^2}\left(\frac{dC}{d\rho}\right)^2 - \frac{1}{4}\frac{1}{C}\frac{d^2C}{d\rho^2} - \frac{1}{4}\frac{d}{d\rho}(A-D) +$$
$$+\frac{1}{4}\frac{(A-D)}{C}\frac{dC}{d\rho} + \frac{1}{8}(A-D)^2 + \frac{1}{2}BC. \tag{23}$$

**8.** Square integrability of $\psi_F(\rho)$, $\psi_G(\rho)$

The important circumstance is that with the transition to the Schrödinger type equations, radial wave functions $\psi_F(\rho)$, $\psi_G(\rho)$ become square integrable in all domains of definition $\rho \in [\rho_+, \infty)$, $\rho \in (0, \rho_-]$, $\rho \in [2\alpha, \infty)$.

$$\psi_F\left(\varepsilon = \frac{\alpha_{em}}{\rho_+}\right)\bigg|_{\rho \to \rho_+} = C_1(\rho - \rho_+)^{1/4}, \tag{24}$$

$$\psi_F\left(\varepsilon = \frac{\alpha_{em}}{\rho_-}\right)\bigg|_{\rho \to \rho_-} = C_2(\rho_- - \rho)^{1/4}, \tag{25}$$

$$\psi_F(\varepsilon = 0)\big|_{\rho \to 2\alpha} = C_3(\rho - 2\alpha)^{1/4}. \tag{26}$$

Wave functions are zero on event horizons $\rho_+, \rho_-, 2\alpha$.

### 3. Energies of stationary bound states of spin half particles in external Schwarzschild, Reissner-Nordström, Kerr and Kerr-Newman fields

The Schwarzschild field:
$$\varepsilon_S = 0. \tag{27}$$

The Reissner-Nordström field:
$$\rho_+ = \alpha + \sqrt{\alpha^2 - \alpha_Q^2}; \quad \rho_- = \alpha - \sqrt{\alpha^2 - \alpha_Q^2},$$
$$\varepsilon_{R-N} = \alpha_{em}/\rho_+, \quad \rho \in [\rho_+, \infty], \tag{28}$$
$$\varepsilon_{R-N} = \alpha_{em}/\rho_-, \quad \rho \in (0, \rho_+]. \tag{29}$$

The Kerr field:
$$\rho_+ = \alpha + \sqrt{\alpha^2 - \alpha_a^2}; \quad \rho_- = \alpha - \sqrt{\alpha^2 - \alpha_a^2},$$
$$\varepsilon_K = \frac{m_\varphi \alpha_a}{\alpha_a^2 + \rho_+^2}, \quad \rho \in [\rho_+, \infty). \tag{30}$$

$$\varepsilon_K = \frac{m_\varphi \alpha_a}{\alpha_a^2 + \rho_-^2}, \quad \rho \in (0, \rho_-]. \tag{31}$$

The Kerr-Newman field:

$$\rho_+ = \alpha + \sqrt{\alpha^2 - \alpha_a^2 - \alpha_Q^2}; \quad \rho_- = \alpha - \sqrt{\alpha^2 - \alpha_a^2 - \alpha_Q^2},$$

$$\varepsilon_{K-N} = \frac{m_\varphi \alpha_a + \alpha_{em} \rho_+}{\rho_+^2 + \alpha_a^2}, \quad \rho \in [\rho_+, \infty), \tag{32}$$

$$\varepsilon_{K-N} = \frac{m_\varphi \alpha_a + \alpha_{em} \rho_-}{\alpha_a^2 + \rho_-^2}, \quad \rho \in (0, \rho_-]. \tag{33}$$

Interval $-1 \leq \varepsilon \leq 1$ is the allowed interval of the particle energy in the bound state.

For each solutions there are square integrable eigenfunctions that are different for different $j, l$ and are solutions of the Schrödinger type equations with effective potentials.

## 4. Schwarzschild and Reissner-Nordström collapsars with bound fermions are candidates for "dark matter" particles

**4.1.** Let us consider a solution for the Schwarzschild field $\varepsilon_S = 0$. If we neglect the gravitational interaction of the uncharged Dirac particles, the atomic-type system of bound half-spin particles is allowable for the Schwarzschild collapsar with mass $M$ and $\varepsilon_S = 0$. Degenerated states with different $m_\varphi$ should be filled considering the Pauli principle. An example is the hydrogen atom with states degenerate over orbital momenta $l$.

The atomic system: nonevaporating Schwarzschild collapsar with uncharged Dirac particles with $\varepsilon_S = 0$ interacts with other objects only by gravitation. Since there are no quantum transitions between the states with different $j, l$, this system does not either emit or absorb light or other kinds of radiation. Such a system can be discovered only by gravitational interaction. Masses of such systems should be chosen with consideration of a requirement of the best compliance with the Standard cosmological model.

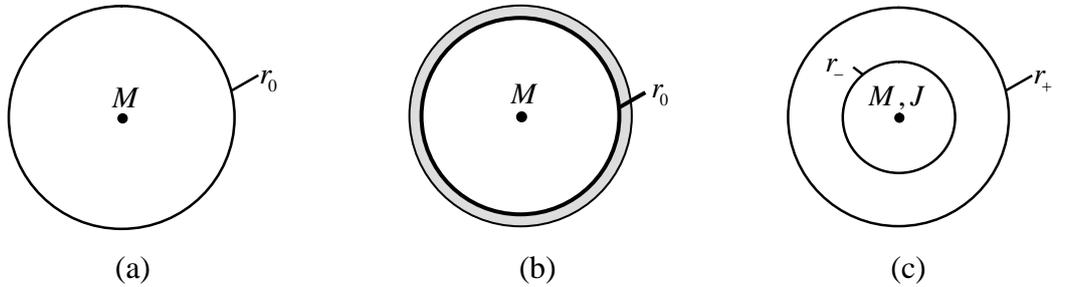

**Fig. 1.** (a) Schwarzschild collapsar, (b) Schwarzschild collapsar with fermions in bound states, (c) Kerr collapsar.

### 4.2. *Solutions for the Reissner-Nordström field (RN)*

Let us consider a solution $\varepsilon = \alpha_{em}/\rho_{-}$. In the case when the formation of a RN collapsar induces creation of an atomic system with bound fermions, which are located near an interior neighborhood of event horizon $\rho_{-}$, and the charge of the RN field source is compensated by the total charge of bound fermions, this atomic system interacts with other objects only gravitationally as seen from the external world. In the absence of quantum transitions between states with different $\kappa$ (or $j,l$), the system does not either radiate or emit light or other kinds of radiation. This system can be discovered only though gravitational interaction.

The system of bound fermions in the RN field with energy $\varepsilon = \alpha_{em}/\rho_{+}$ can be considered as the second atomic system. In this case, fermions with an overwhelming probability are located near the outer neighborhood $\rho_{+}$, and at compensation of charge of the RN field source by the total charge of bound fermions, this atomic system interacts with other external objects only by gravitation. Like in the first case, the atomic system does not either emit or radiate light or other kinds of radiation. The charge of the RN field source can be detected in this atomic system only if we "knock out" some of the fermions from their orbits by an external action.

Some other atomic systems partially with the energy of bound fermions $\varepsilon = \alpha_{em}/\rho_{+}$, and partially with $\varepsilon = \alpha_{em}/\rho_{-}$ are possible candidates for dark matter particles. In the presence of rotation (Kerr field, Kerr-Newman field) the situation does not vary qualitatively.

Masses of the considered systems should be chosen taking into consideration the requirement of the best compliance with the Standard Cosmological Model.

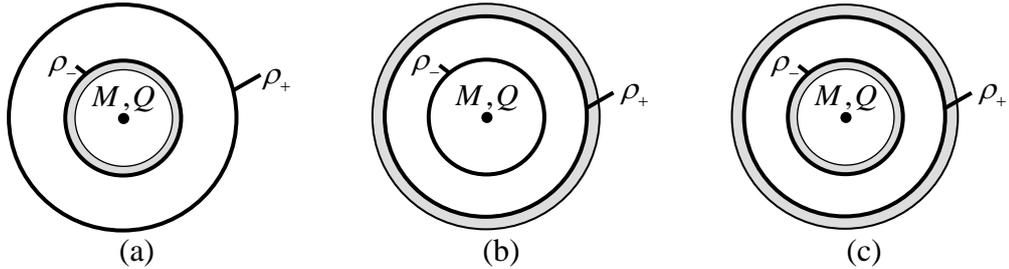

(a)          (b)          (c)

**Fig. 2.** (a, b, c) Reissner-Nordström collapsars with fermions in bound states.

## 5. Conclusions

After consideration of solutions of the Schrödinger type equations with effective potentials in the quantum mechanics of fermion motion in the classical Schwarzschild and Reissner-Nordström fields, the following results are obtained:

(1) In the presence of event horizon, there are regular solutions with energies $\varepsilon = 0$ (Schwarzschild field), $\varepsilon = \alpha_{em}/\rho_{+}$, $\varepsilon = \alpha_{em}/\rho_{-}$ (RN field). These solutions are the stationary bound states of charged and uncharged fermions with square integrable wave functions and with domains of definition $\rho \in [2\alpha, \infty), \rho \in [\rho_+, \infty), \rho \in (0, \rho_-]$. Wave functions weakly depend on $j, l$ and vanish on the event horizons. Fermions in bound states with overwhelming probability are located near the event horizons. Maxima of probability densities are fractions to units of the fermion Compton wavelength away from the event horizons.

(2) Electrically neutral atomic systems with the particular number of fermions in degenerate bound states with $\varepsilon = 0$ (Schwarzschild field), and $\varepsilon = \alpha_{em}/\rho_+$, $\varepsilon = \alpha_{em}/\rho_-$ (RN field) can be considered in the standard cosmological model as particles of dark matter. Atomic systems of such type do not either absorb or emit light or other kinds of radiation and interact with surrounding medium only by gravitation.

(3) In spite of the removal of the degeneracy with respect to the magnetic quantum number $m_\varphi$, atomic systems with bound stationary states of fermions with $\varepsilon = \dfrac{m_\varphi \alpha_a}{\alpha_a^2 + \rho_+^2}$,

$\varepsilon = \dfrac{m_\varphi \alpha_a}{\alpha_a^2 + \rho_-^2}$ (Kerr field); and $\varepsilon = \dfrac{m_\varphi \alpha_a + \alpha_{em}\rho_+}{\alpha_a^2 + \rho_+^2}$, $\varepsilon = \dfrac{m_\varphi \alpha_a + \alpha_{em}\rho_-}{\alpha_a^2 + \rho_-^2}$ (Kerr-Newman field) can be considered as dark matter particles under certain conditions.

Regular solutions for bound states with fermion energies of $\varepsilon = 0$ (Schwarzschild metric) and $\varepsilon = \alpha_{em}/\rho_+$, $\varepsilon = \alpha_{em}/\rho_-$ (Reissner-Nordström metric) are obtained using the Schrödinger type equation with effective potentials. The wave function is associated with one of radial functions from the Dirac equation by a nonunitary transformation. As a result, wave functions of the Schrödinger type equation for stationary bound states, unlike radial functions of the Dirac equation, become square integrable in the neighborhoods of event horizons $\rho_+, \rho_-$. The Schrödinger type equation can be also obtained by squaring the covariant Dirac-Fock equation in non-Euclidean space-time with transition from a bispinor to a spinor wave function and with corresponding non-unitary transformation. The covariant second-order equation for fermions moving in external electromagnetic fields was proposed by P.Dirac for the Minkowski flat space-time in the 1930s.

Our consideration shows that using the relativistic second-order equation expends our opportunities to obtain the regular solutions of the quantum mechanics of half-spin particle motion in external gravitational fields.

**References**

[1] V. P. Neznamov and I. I. Safronov. VANT, Ser. Theor. Appl. Phys. **4**, 9-24 (2016).
[2] V. P. Neznamov, I. I. Safronov, and V.E.Shemarulin. VANT, Ser. Theor. Appl. Phys. **2**, 12-40 (2017).